# Structural analysis and transport properties of [010]-tilt grain boundaries in Fe(Se,Te)

Kazumasa Iida[a,b]* ,Yoshihiro Yamauchi[c], Takafumi Hatano[b,c], Kai Walter[d], Bernhard Holzapfel[d], Jens Hänisch[d], Zimeng Guo[b,e], Hongye Gao[f], Haoshan Shi[g], Shinnosuke Tokuta[b,h], Satoshi Hata[b,e,f,g], Akiyasu Yamamoto[b,h] and Hiroshi Ikuta[c,i]

[a]College of Industrial Technology, Nihon University, Narashino, Chiba, Japan;

[b]JST CREST, Kawaguchi, Saitama, Japan;

[c]Department of Materials Physics, Nagoya University, Nagoya, Japan;

[d]Institute for Technical Physics, Karlsruhe Institute of Technology, Eggenstein-Leopoldshafen, Germany;

[e]Department of Advanced Materials Science and Engineering, Kyushu University, Kasuga, Fukuoka, Japan;

[f]The Ultramicroscopy Research Center, Kyushu University, Fukuoka, Japan;

[g]Interdisciplinary Graduate School of Engineering Sciences, Kyushu University, Kasuga, Fukuoka, Japan;

[h]Department of Applied Physics, Tokyo University of Agriculture and Technology, Koganei, Tokyo, Japan;

[i]Research Center for Crystalline Materials Engineering, Nagoya University, artment, University, City, Country; [b]Department, University, City, Country

# Structural analysis and transport properties of [010]-tilt grain boundaries in Fe(Se,Te)

Understanding the nature of grain boundaries is a prerequisite for fabricating high-performance superconducting bulks and wires. For iron-based superconductors [e.g. Ba(Fe,Co)$_2$As$_2$, Fe(Se,Te),and NdFeAs(O,F)], the dependence of the critical current density $J_c$ on misorientation angle ($\theta_{GB}$)has been explored on [001]-tilt grain boundaries, but no data for other types of orientations have been reported. Here, we report on the structural and transport properties of Fe(Se,Te) grown onCeO$_2$-buffered symmetric [010]-tilt roof-type SrTiO$_3$ bicrystal substrates by pulsed laser deposition. X-ray diffraction and transmission electron microscopy revealed that $\theta_{GB}$ of Fe(Se,Te) was smaller whereas $\theta_{GB}$ of CeO$_2$ was larger than that of the substrate. The difference in $\theta_{GB}$ betweenthe CeO$_2$ buffer layer and the substrate is getting larger with increasing $\theta_{GB}$. For $\theta_{GB} \geq 24^o$ of the substrates, $\theta_{GB}$ of Fe(Se,Te) was zero, whereas $\theta_{GB}$ of CeO$_2$ was continuously increasing. The inclined growth of CeO$_2$ can be explained by the geometrical coherency model. The $c$-axis growth of Fe(Se,Te) for $\theta_{GB} \geq 24^o$ of the substrates is due to the domain matching epitaxy on (221) planesof CeO$_2$. Electrical transport measurements confirmed no reduction of inter-grain $J_c$ for $\theta_{GB} \leq 9^o$, indicative of strong coupling between the grains.

## 1. Introduction

Grain boundaries (GBs) are interfaces between crystalline grains at which the crystallographic orientation abruptly changes. Microscopically, the overlap of the wave functions is perturbed by GBs, leading to a change in the electronic structure. The electronic structure is also affected by local strain and dislocations in and around the GB. Hence, physical properties across GBs are expectedly altered, and understanding the nature of GBs is therefore an important step for further improvement of the functionalities of materials. Polycrystalline samples contain many types of GBs, which complicates the investigations of specific GBs. To understand the nature of such a specific GB, they have to be fabricated artificially. For instance, the attempt at realizing artificial GBs in silicon ingots has been reported recently [1]. For high-temperature superconductors (HTS, e.g.

YBa$_2$Cu$_3$O$_{7-\delta}$, YBCO) as well as iron-based superconductors (IBSs), thin films containing a well-defined single GB have been fabricated, since the critical current would be too large to evaluate by electrical transport measurements on bulk samples. In this case, superconducting thin films have been grown biaxially on bicrystal substrates, which consist of two single crystals having a, usually common symmetric, rotation along [001] that are joined by a solid-state reaction [2]. After growth, the electrical transport properties across the GB are investigated as a function of misorientation angle. Such experiments are recognized as a powerful method for understanding the GB properties of HTS, for reviews see [3,4].

For cuprates, not only GBs with in-plane misorientation ([001]-tilt GB) but also with out-of-plane misorientation ([010]-tilt GB) as well as [100]- and [001]-twist GBs have been realized [5,6]. The inter-grain $J_c$ across [001]-tilt GBs was shown to decrease exponentially above a $\theta_{GB}$ around 3º~5º [2,3,5]. This angle is defined as the critical angle $\theta_c$. Similar to the [001]-tilt GBs, the inter-grain $J_c$ reduced significantly at the [100]-twist type GBs. On the other hand, the inter-grain $J_c$ of [001]-twist GBs for Bi$_2$Sr$_2$CaCu$_2$O$_{8+\delta}$ was unaltered regardless of misorientation angle [7]. For YBCO, the inter-grain $J_c$ of [010]-tilt GBs was almost constant even for $\theta_{GB}$ = 8º [5,6], indicating that $\theta_c$ can depend on the type of GB.

For Ba(Fe,Co)$_2$As$_2$ [8], Fe(Se,Te) [9,10], and NdFeAs(O,F) [11,12], only [001]-tilt GBs have been investigated so far. The common feature of those IBSs is that $\theta_c$ is around 9º, which is 2–3 times larger than for YBCO of the same type of GB. Additionally, the inter-grain $J_c$ stayed constant in the range 15º≤ $\theta_{GB}$ ≤45º, whereas for YBCO it decreases further exponentially with $\theta_{GB}$. These prominent features of GBs in IBSs may originate from their $s\pm$ wave symmetry. However, no data for other types of orientations have been reported. Hence, it is interesting how $J_c$ is affected by [010]-tilt as well as twist GBs. To address this issue, we have fabricated Fe(Se,Te) thin films on symmetric [010]-tilt roof-type SrTiO$_3$ bicrystals with $\theta_{GB}$ up to 30º and investigated the structural and transport properties.

We have selected Fe(Se,Te), since it has the simplest crystal structure among IBSs. Hence, it is considered easy to extract the factors governing the superconducting properties. However, growing Fe(Se,Te) thin films with good superconducting properties is not easy due to the excess Fe, which localizes conducting carriers, leading to a lower

$J_c$ [13]. In fact, as-grown films under our growth conditions contain excess Fe.

In this paper, we firstly optimize the post-annealing conditions for Fe(Se,Te) to remove excess Fe. Then, Fe(Se,Te) bicrystal films are fabricated by employing the optimized post-annealing condition, followed by structural and electrical transport characterizations.

## 2. Experiment

$CeO_2$ was grown on $SrTiO_3(001)$ (K&R Creation Co., Ltd. Japan) in $pO_2 = 1$ Pa at 600 ºC by pulsed laser deposition (PLD), where a commercially available $CeO_2$ sintered target (Toshima Manufacturing Co.,Ltd. Japan) was ablated by a KrF excimer laser (COMPex 102F, Coherent Inc., USA) (wavelength $\lambda = 248$ nm) with 1 Hz. An energy density of ~1.2 J/cm$^2$ at the target surface was employed. A total pulse number of 1320 yielded a 30 nm-thick $CeO_2$ film confirmed by X-ray reflectivity measurements (Supplementary Figure S1). After deposition, the $CeO_2$-buffered $SrTiO_3$ substrates were transferred to the UHV chamber (base pressure ~1×10$^{-7}$ Pa) for deposition of Fe(Se,Te) without exposing them to air. The Fe(Se,Te) target with nominal composition Fe:Se:Te = 1:0.5:0.5 was prepared by spark plasma sintering [14]. The precursor powders were mechanically alloyed prior to the sintering [15]. The nominal $FeSe_{0.5}Te_{0.5}$ films were also grown on $CeO_2$-buffered [010]-tilt roof-type $SrTiO_3$ bicrystal substrates ($8º \leq \theta_{GB}^{STO} \leq 30º$, Furuuchi Chemical Co., Japan) at 300 ºC and with 5 Hz laser repetition rate. The energy density of the laser was the same as for the $CeO_2$ deposition. A pulse number of 7500 yielded a 135–155 nm-thick $FeSe_{0.5}Te_{0.5}$ layer, which is the optimum thickness for achieving a high $T_c$ [16,17].

Post annealing has been conducted by referring to [18,19]. The samples were again transferred to the $CeO_2$ deposition chamber after growth of $FeSe_{0.5}Te_{0.5}$ followed by annealing at 100 ºC $\leq T_{anneal} \leq 350$ ºC in a fixed $pO_2$ of 1 Pa. The dwell time at the maximum $T_{anneal}$ for each experimental run was fixed at 10 min.

Structural properties of the films were characterized by X-ray diffraction (XRD, RINT2000 and ULTIMA IV, RIGAKU, Japan) using Cu Kα radiation and transmission electron microscopy (TEM). The [001] directions of both $FeSe_{0.5}Te_{0.5}$ and $CeO_2$ are expected to be away from the substrate normal by $\theta_{GB}^{STO}/2$ when $FeSe_{0.5}Te_{0.5}$ is grown on

CeO$_2$-buffered symmetric [010]-tilt SrTiO$_3$ bicrystal substrates having $\theta_{GB}^{STO}$. Hence, the growth angles (i.e. offset angles) for FeSe$_{0.5}$Te$_{0.5}$ and CeO$_2$ were determined by $\omega$-scans, where the angle $2\theta$ was fixed to the 002 reflections of each layer. TEM was performed on a cross-sectional foil sample covering the grain boundary. The foil sample was made by focused ion beam (FIB) in a scanning electron microscope (SEM) called Helios Hydra CX (Thermo Fisher Sci., USA). The scanning TEM (STEM) observations were carried out for high-resolution microstructural analyses by a TEM called Titan Cubed G2 (Thermo Fisher Sci., USA). In order to accurately assess the grain boundary angle in each layer, the automated crystal orientation mapping (ACOM) technique in a TEM called ARM-200F (JEOL Ltd., Japan) was performed by using ASTAR device (NanoMEGAS, Belgium) with a spatial resolution of 4 nm and an acceleration voltage of 200 kV. Details of the ACOM in TEM are described in refs. [20,21].

After structural characterization, electrical transport properties were measured using a 4-probe method on micro-bridges fabricated by laser cutting (UV-MK-kit, Kokyo, Inc., Japan). The bridges of 100 μm width had a length of 2 mm for inter-grain measurements, and 1 mm for intra-grain measurements, respectively. The superconducting transition temperature ($T_{c,90}$) was defined as a 10% drop of the normal state resistance $R_n$, at which the resistance deviated from the linear fit to the normal state in the vicinity of the superconducting transition. $J_c$ was determined by an electrical field criterion of 1 μV/cm.

## 3. Results and discussion

### 3.1. Removal of excess Fe

The as-grown FeSe$_{0.5}$Te$_{0.5}$ films on CeO$_2$-buffered ordinary SrTiO$_3$(001) substrates contained excess Fe, inferred from a resistance upturn before the superconducting transition (Figure 1(a)). This upturn is due to the charge carrier localization by excess Fe in Fe(Se,Te) [13]. Figure 1(a) shows the normalized resistance curves of the FeSe$_{0.5}$Te$_{0.5}$ thin films after post-annealing. The resistance upturn was gradually suppressed with increasing $T_{anneal}$. At 200 ºC$\leq T_{anneal} \leq$220 ºC, the upturn disappeared. Simultaneously, the superconducting transition temperature $T_{c,90}$ increased with $T_{anneal}$ and reached a maximum around 15 K at $T_{anneal}$ =200 ºC (Figure 1(b)). Further increasing $T_{anneal}$ reduced $T_{c,90}$. For $T_{anneal}$ > 300 ºC, superconductivity disappeared completely. Additionally, the resistance curve for the film annealed at 300 ºC showed semiconducting behavior. Figure

1(c) shows the XRD patterns of FeSe$_{0.5}$Te$_{0.5}$ annealed at various temperatures. In the XRD $2\theta/\omega$ scans, no appreciable differences between the as-grown film and the film annealed at 200 ºC were observed. On the other hand, significant shifts of the 00l reflections toward higher $2\theta$ values were observed for the film annealed at 300 ºC, indicative of a decrease in c-axis length. This is mainly due to the loss of Te, since severe annealing conditions may terminate the Fe-Te bonds leading to a loss in Te [22] and the *c*-axis length is decreasing with decreasing Te content in FeSe$_{0.5}$Te$_{0.5}$ single crystal [23]. When the film was annealed at 350 ºC, further shifting of the 00$l$ peaks together with peaks originating from impurities was recognized. In fact, the $c$-axis length significantly reduced at $T_{anneal}$ ≥ 300 ºC (Figure 1(d)), whereas the $c$-axis length of the superconducting films located between 6.0 Å and 6.1 Å. From those results, the optimum post-annealing temperature was determined as 200 ºC.

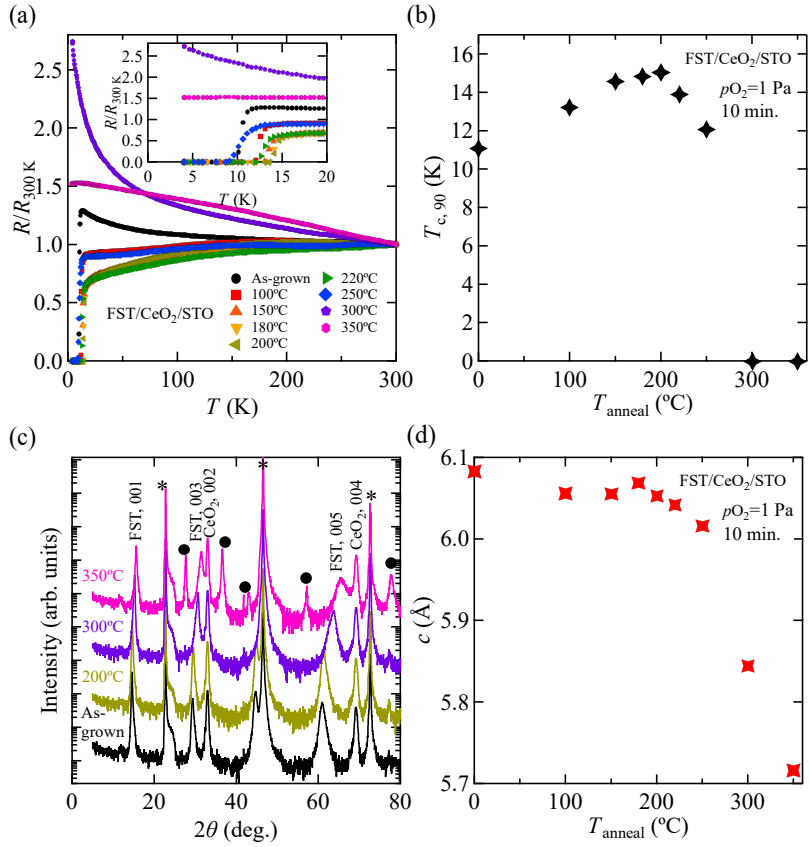

**Figure 1.** (a) The resistance curves of the as-grown FeSe$_{0.5}$Te$_{0.5}$ (FST) and the FST films annealed at various temperatures normalized to the value at 300 K. Inset shows a magnified image of (a) around the superconducting transition. STO represents the SrTiO$_3$ substrate. (b) The transition temperature $T_{c,90}$ as a function of the annealing temperature $T_{anneal}$. The maximum $T_{c,90}$ around 15 K was observed at $T_{anneal}$ =200ºC. (c) The XRD patterns of the FeSe$_{0.5}$Te$_{0.5}$ thin films annealed at 200ºC, 300ºC, and 350ºC. For comparison, the data for the as-grown film is also shown. Beyond the optimum $T_{anneal}$, the 00l diffraction peaks shifted to higher angles. For $T_{anneal}$ =350ºC, some diffraction peaks marked as "•" other than FST and CeO$_2$ were observed. The peaks marked as "∗" originate from SrTiO$_3$. (d) The *c*-axis length as a function of the annealing temperature $T_{anneal}$. The *c*-axis length of the superconducting films was located between 6.0 Å and 6.1 Å.

The post-annealing conditions in this study differed from the ones reported by Zhang *et al*. [18] with respect to $pO_2$, annealing temperature and dwell time, which were there 100 mbar (~13.3 Pa), 90 ºC and 1~2 h. The annealing temperature of 200 ºC in our case is almost the double of Zhang's study, whereas our dwell time is shorter. In our study, the resistance upturn was suppressed even at $T_{anneal}$ =100 ºC. Hence, it may be possible to remove more Fe with further increasing the holding time. The post-annealing reported by Zhang et al. not only led to removal of excess Fe but also to a significant enhancement of critical currents, although $T_c$ was slightly reduced. Post-annealing at low temperatures may indeed be used to tune the properties of superconducting films further, such as critical current properties of REBCO films [24]. Nevertheless, in the following, the FeSe$_{0.5}$Te$_{0.5}$ films on CeO$_2$-buffered [010]-tilt SrTiO$_3$ bicrystal substrates were post-annealed at 200 ºC for 10 min in 1 Pa of oxygen.

### 3.2. Structural analyses

Figure 2(a) exhibits the XRD 2$\theta$/$\omega$ patterns of FeSe$_{0.5}$Te$_{0.5}$ grown on CeO$_2$-buffered [010]-tilt SrTiO$_3$ bicrystal substrates with various misorientation angles. The film for $\theta_{GB}^{STO}$= 0º was grown on an ordinary SrTiO3 (001) substrate. The angle $\theta_{GB}^{FST}$ shown in the panel indicates the measured offset angle of FeSe$_{0.5}$Te$_{0.5}$ multiplied by two [i.e. the actual misorientation angle of FeSe$_{0.5}$Te$_{0.5}$], and the angle in parenthesis is the misorientation angle of the SrTiO$_3$ bicrystals ($\theta_{GB}^{STO}$). For $\theta_{GB}^{STO}$= 0º, the 00$l$ reflections of FeSe$_{0.5}$Te$_{0.5}$ and CeO$_2$ together with SrTiO$_3$ were observed. Additionally, the 101 reflection of the $\phi$ scan showed a fourfold symmetry [Supplementary Figure S2(a)], which proves the phase-pure and epitaxial growth of FeSe$_{0.5}$Te$_{0.5}$.

On the other hand, almost only the 00l reflections of FeSe$_{0.5}$Te$_{0.5}$ were observed for $\theta_{GB}^{STO}$ > 0º, indicating that the offset angle of FeSe$_{0.5}$Te$_{0.5}$ differs from those of CeO$_2$ and SrTiO$_3$. In fact, the respective misorientation angles of FeSe$_{0.5}$Te$_{0.5}$ and CeO$_2$ are different from each other and from those of the SrTiO$_3$ bicrystals (Figure 2(b)). As can be seen, the actual misorientation angle of CeO$_2$ ($\theta_{GB}^{CeO2}$) is getting larger than $\theta_{GB}^{STO}$, whereas $\theta_{GB}^{FST}$ is always smaller than $\theta_{GB}^{STO}$. A similar effect was observed in FeSe$_{0.5}$Te$_{0.5}$ thin films on vicinal CaF$_2$ substrates deposited at 260 °C [28]. For $\theta_{GB}^{STO} \geq 24°$, $\theta_{GB}^{FST}$ was zero, indicating the absence of a GB in FeSe$_{0.5}$Te$_{0.5}$. These observations can be explained by the geometrical coherency model [25–27], according to which $\theta_{GB}^{CeO2} - \theta_{GB}^{STO} = \Delta\theta_{GB1}$ and $\theta_{GB}^{FST} - \theta_{GB}^{CeO2} = \Delta\theta_{GB2}$ can be calculated

$$\frac{\Delta\theta_{GB1}}{2} = \left| \tan^{-1}\left( \frac{d_{STO} - d_{CeO2}}{d_{STO}} \tan\frac{\theta_{GB}^{STO}}{2} \right) \right| \quad (1)$$

$$\frac{\Delta\theta_{GB2}}{2} = \left| \tan^{-1}\left( \frac{d_{CeO2} - d_{FST}}{d_{CeO2}} \tan\frac{\theta_{GB}^{CeO2}}{2} \right) \right| \quad (2)$$

where $d_{STO}$, $d_{CeO2}$, and $d_{FST}$ are the out-of-plane, monolayer step height of SrTiO$_3$ (3.91 Å), CeO$_2$ (5.41 Å), and FeSe$_{0.5}$Te$_{0.5}$ (5.96 Å), respectively. The direction of the tilt of [001] CeO$_2$ from [001] SrTiO$_3$ is away from the substrate normal, because $d_{CeO2} > d_{STO}$. Similarly, the direction of the tilt of [001] FeSe$_{0.5}$Te$_{0.5}$ from [001] CeO$_2$ is away from the substrate normal. For $\theta_{GB}^{STO} = 30°$, $\Delta\theta_{GB1}/2$ is calculated as 5.9°, which is close to the measured angle from the STEM image shown in Figure 2(c). The grain boundary angles of CeO$_2$ ($\theta_{GB}^{CeO2}$) lie on the calculated (dotted blue) line (Figure 2(b)), indicating that the geometrical coherency model is valid. However, this model seems not to be valid for FeSe$_{0.5}$Te$_{0.5}$/CeO$_2$, since the experimental data did not lie on the dashed red line calculated from the model. The vicinal angles of FeSe$_{0.5}$Te$_{0.5}$ grown on off-cut CaF$_2$ substrates at 260 ºC deviated similarly from the calculation (Supplementary Figure S3). This may be due to the low growth temperature, leading to a low surface mobility of atoms [27]. In fact, the vicinal angles of FeSe$_{0.5}$Te$_{0.5}$ grown at a higher temperature of 400 ºC were almost identical to those of the CaF$_2$ substrates (Figure S3). Possibly, film surfaces and CaF$_2$ at low temperatures do not have well defined terraces needed for the geometry coherency mechanism. Finally, for a proper analysis, the lattice parameters at growth temperature should be considered, which we omitted here for our estimates.

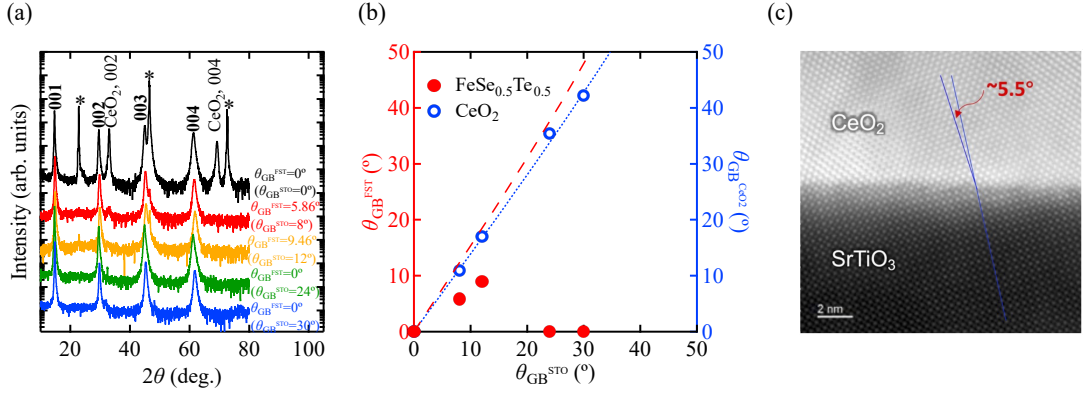

**Figure 2.** (a) The XRD pattern of the FeSe$_{0.5}$Te$_{0.5}$ thin films grown on CeO$_2$-buffered [010]-tilt symmetric SrTiO$_3$ (STO) bicrystal substrates having various grain boundary angles $\theta_{GB}^{STO}$. Here, the angle $\theta_{GB}^{FST}$ corresponds to twice the offset angle of FeSe$_{0.5}$Te$_{0.5}$. The peaks marked as "*" originate from SrTiO$_3$. Because of the different offset angles between FeSe$_{0.5}$Te$_{0.5}$ and CeO$_2$ as well as FeSe$_{0.5}$Te$_{0.5}$ and SrTiO$_3$, almost only the 00$l$ peaks from FeSe$_{0.5}$Te$_{0.5}$ were observed. (b) The $\theta_{GB}^{FST}$ for FeSe$_{0.5}$Te$_{0.5}$ (closed symbol) and $\theta_{GB}^{CeO2}$ for CeO$_2$ (open symbol) as a function of $\theta_{GB}^{STO}$. The dashed red and dotted blue lines are calculations using the geometrical coherency model [25–27]. (c) The atomic resolution HAADF-STEM image of the interface between CeO$_2$ and SrTiO$_3$ having a $\theta_{GB}^{STO}$ = 30°. The calculated value of $(\theta_{GB}^{CeO2}, \theta_{GB}^{STO})/2$ is 5.9°, which is close to the measured value of ~5.5°.

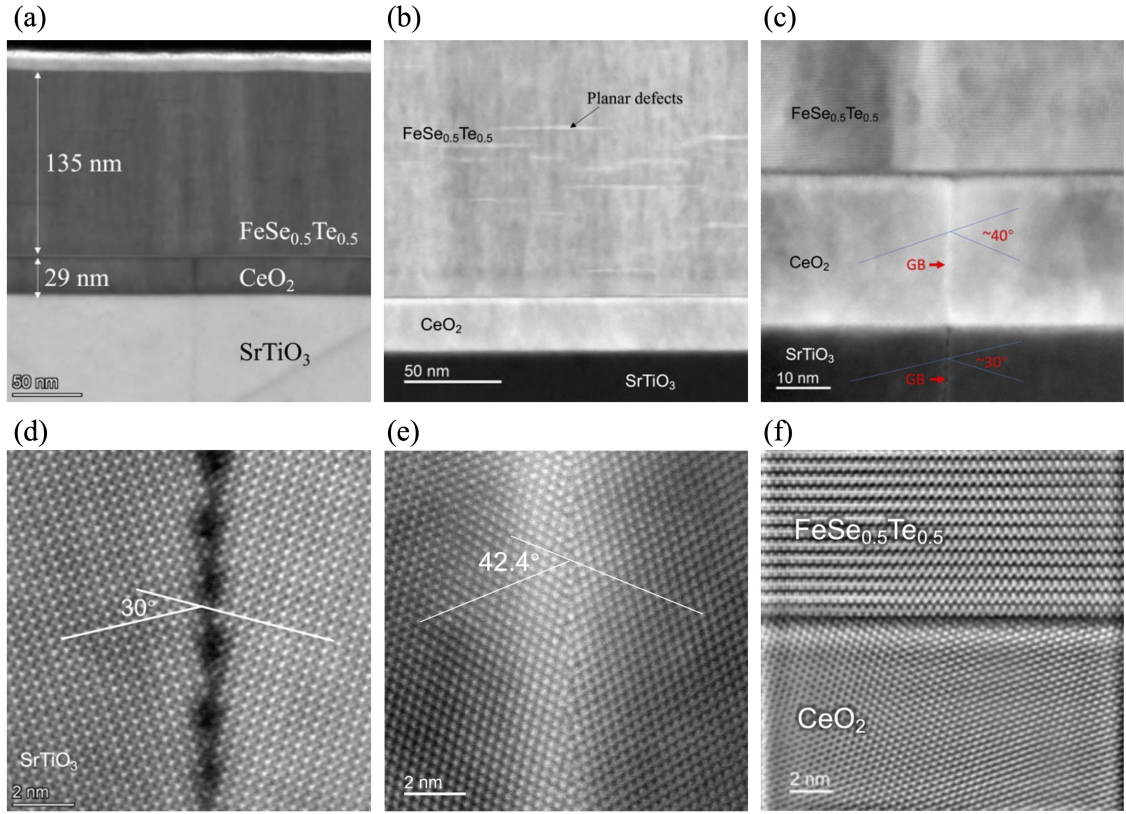

**Figure 3.** Microstructure of the FeSe$_{0.5}$Te$_{0.5}$/CeO$_2$ sample grown on the 30º [010]-tilt symmetric SrTiO$_3$ bicrystal substrate. (a) Cross-sectional view near the GB acquired by ADF-STEM. (b) ADF-STEM image taken away from GB. Planar defects shown by a black arow are visible. (c) Magnified image of (a). The GB is absent in FeSe$_{0.5}$Te$_{0.5}$. Atomic-resolution HAADF-STEM image of the GBs in SrTiO$_3$ (d) and CeO$_2$ (e). The GB angle in CeO$_2$, $\theta_{GB}^{CeO2}$, is 42.4º, consistent with the value by XRD. (f) Atomic-resolution image of the interface between CeO$_2$ and FeSe$_{0.5}$Te$_{0.5}$, which was clean and without reaction layer.

Figure 3(a) shows the cross-sectional view of FeSe$_{0.5}$Te$_{0.5}$ grown on the CeO$_2$-buffered SrTiO$_3$ bicrystal with $\theta_{GB}^{STO} = 30$º. The respective layer thicknesses of FeSe$_{0.5}$Te$_{0.5}$ and

$CeO_2$ were 135 nm and 30 nm. The film contained planar defects with a thickness of ~1.5 nm along the ab-plane, Figure 3(b). Atomic-resolution images of $SrTiO_3$ and $CeO_2$ buffer layer around the GB confirmed that the respective GB angles are $\theta_{GB}^{STO} = 30°$ and $\theta_{GB}^{CeO2} = 42.4°$ (Figures 3(d,e)). Those values are consistent with the ones evaluated by XRD measurements. Figure 3(c) confirms the presence of a GB in the $CeO_2$ buffer layer, whereas no visible GB was present in the $FeSe_{0.5}Te_{0.5}$ layer as stated above. Additionally, the $FeSe_{0.5}Te_{0.5}$ layer grew biaxially textured as shown in Figure 3(f). The in-plane texture was also confirmed by the $\phi$ scan of the 101 reflection [Supplementary Figure S2(b)]. According to the geometric considerations based on the TEM observation, the epitaxial relation $(001)[100]FeSe_{0.5}Te_{0.5} \parallel (114)[22\bar{1}]CeO_2$ is realized as a domain growth [29]. In fact, a domain wall structure was observed in the $FeSe_{0.5}Te_{0.5}$ film along [010], i.e. across the GB for $\theta_{GB}^{STO} = 30°$, and their average width was $32 \pm 12$ nm (Figures 4(a,d)). Note that such a structure has not been observed in the $FeSe_{0.5}Te_{0.5}$ film grown on $CeO_2$-buffered single-crystal $SrTiO_3$ substrate (Supplementary Figure S4). The relation $(001)[100]FeSe_{0.5}Te_{0.5} \parallel (114)[22\bar{1}]CeO_2$ also holds for $\theta_{GB}^{STO} = 24°$. Due to the extinction rule, the diffraction peak arising from the 114 reflection of $CeO_2$ could not be observed in XRD pattern.

The domain growth is expressed by the following indices, $C_{h,k,l}^{m,n,o}$, where ($h \times k \times l$) lattice of the $CeO_2$ buffer layer and ($m \times n \times o$) lattice of the Fe(Se,Te), refer to [29]. The respective indices are $C_{1,\bar{1},0}^{2,0,0}$ for along the GB and $C_{2,2,1}^{4,0,0}$ for across the GB. However, the most probable index for the latter is $C_{4,4,\bar{2}}^{9,0,0}$ since the domain misfit ($\varepsilon_d$) expressed by Equation (3) is smaller, as shown in Table 1.

$$\varepsilon_d = 2 \frac{\sqrt{m^2 + n^2 + o^2} a_{FST} - \sqrt{h^2 + k^2 + l^2} a_{CeO2}}{\sqrt{m^2 + n^2 + o^2} a_{FST} + \sqrt{h^2 + k^2 + l^2} a_{CeO2}} \qquad (3)$$

Additionally, the domain width $9 \times a_{FST} = 34$ nm ($a_{FST}$: in-plane lattice parameter of $FeSe_{0.5}Te_{0.5}$) corresponds well to the average domain width of 32 nm observed in ACOM, and the opposite mismatch compared to the $FeSe_{0.5}Te_{0.5}$ (100) direction may slightly lower the total energy. The $\varepsilon_d$ of $C_{1,\bar{1},0}^{2,0,0}$ is smaller than that of $C_{4,4,\bar{2}}^{9,0,0}$, which is reflected in the full width at half maximum values ($\Delta\omega$) of the 00l rocking curves [Supplementary Figure S2(h), (k)].

**Table 1.** The domain indices $C_{h,k,l}^{m,n,o}$ and the corresponding domain mismatch calculated from equation (3).

| $C_{h,k,l}^{m,n,o}$ | $\varepsilon_d$ (%) |
|---|---|
| $C_{1,\bar{1},0}^{2,0,0}$ | -0.93 |
| $C_{2,2,\bar{1}}^{4,0,0}$ | -6.82 |
| $C_{4,4,\bar{2}}^{9,0,0}$ | 4.96 |

As can be seen, the $\Delta\omega$ for the $[1\bar{1}0]$ (along the GB, denoted as 'L' in Figure S2) is smaller than that for the $[22\bar{1}]$ (across the GB, denoted as 'T' in Figure S2). For cubic lattices, the $\sum$ ⬚ value of symmetrical GB is expressed by the sum of the squares of the Miller indices [30]. In our experimental results, a $\sum$ 9 [110]/{221} GB with an ideal GB angle of 38.9º has formed in CeO$_2$ on both 24º and 30º substrates with sufficiently close real GB angles of 35.4º and 42.2º, respectively. Unlike other GBs (e.g. $\sum$ 11 [110]/{332}), the $\sum$ 9 [110]/{221} GB is, together with the twin $\sum$ 3 [110]/{111} (not observed here), the most stable structure in CeO$_2$ [31].

Due to the difference between the ideal and real GB angle in CeO$_2$ (3.5º for $\theta_{GB}^{STO} = 24º$ and 3.3º for $\theta_{GB}^{STO} = 30º$, respectively), the (114) planes on either side are tilted by half of this difference, and a GB angle of ,3.5º should be expected in the FeSe$_{0.5}$Te$_{0.5}$ films, which however is not observed in Figure 4(c). From Figures 4(c,d), the respective misorientation angles between domains were within 4º and 3º for in-plane and out-of-plane. The artificial GB (or rather the two sides of the bicrystal) may be still recognized as a more macroscopic shift of the base line (average) misorientation (with respect to a common starting point) of ,0.7º out-of-plane and ,0.9º in-plane, which, however, is well within the range of domain-to-domain misorientations. Two more effects may explain that. First, the geometry coherency growth may happen again, now on (114) instead of (001), and since the *c*-axis of FeSe$_{0.5}$Te$_{0.5}$ is shorter than the single-layer distance in (114) direction in CeO$_2$, the *c*-axis will tend towards the substrate normal, although just about negligible , 0.1º. Secondly, since FeSe$_{0.5}$Te$_{0.5}$ single crystals typically grow in *ab*-oriented platelets, the surface energy of (001) may be concluded to be by far the lowest one. Hence, the system tends to adjust (001) parallel to the surface.

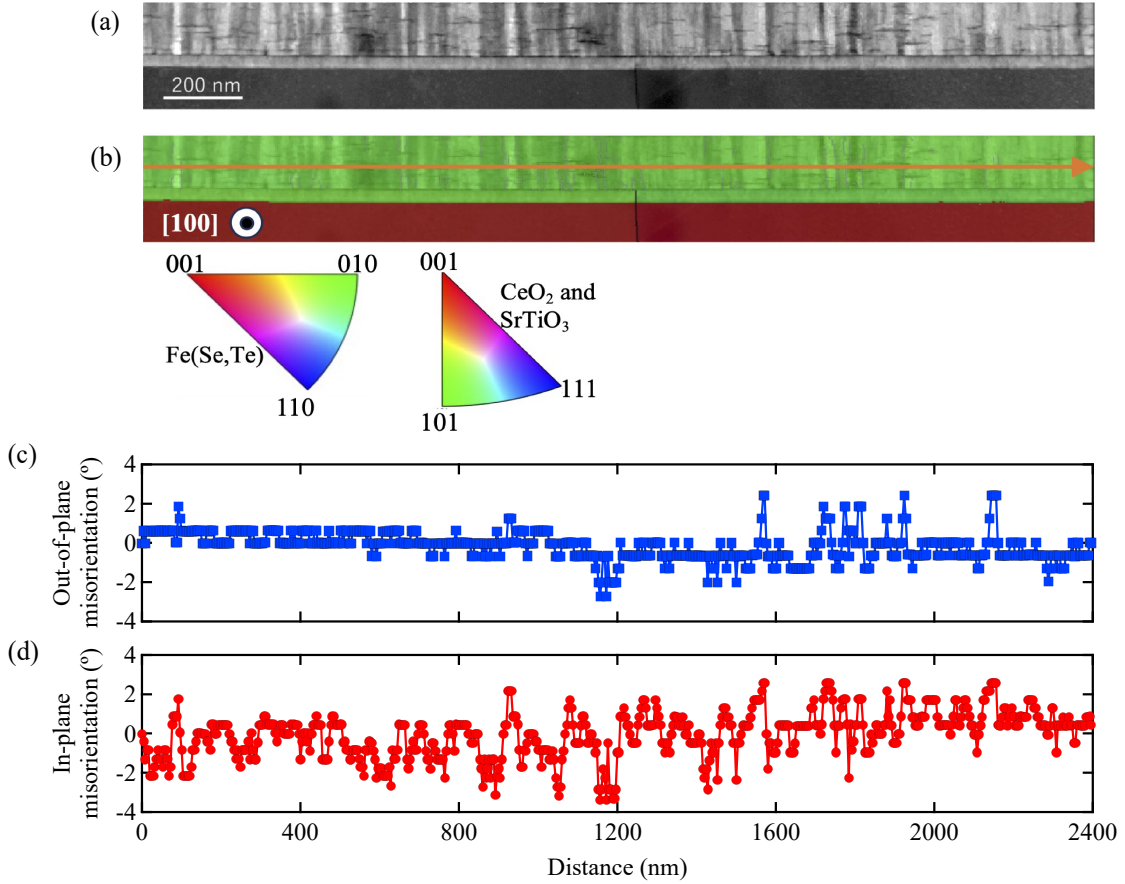

**Figure 4.** Atomic crystal orientation mapping of $FeSe_{0.5}Te_{0.5}$ grown on $CeO_2$-buffered $SrTiO_3$ with $\theta_{GB}^{STO}$=30º by scanning precession diffraction. (b) Inverse pole figure map. (c) Out-of-plane and (d) in-plane misorientation profiles from the first point along the orange line shown in (b).

For the low-angle GBs, a similar combination of special GB in $CeO_2$ ($\sum 33$ [110]/ {441} or $\sum 51$ [110]/{551} may be candidates), geometry coherency on the relevant, vicinal planes [(118) or (11$\underline{10}$) for the above-mentioned GBs], and surface energy reduction may explain the FST GB angles being lower than expected.

### 3.3. Transport properties

The temperature dependence of the resistivity ρ of the inter- and intra-grain bridges is shown in Figure 5. Although $FeSe_{0.5}Te_{0.5}$ GBs were absent for $\theta_{GB}^{STO}$ = 24º and 30º, the data for inter-grain were acquired from the bridges located on the GB of $CeO_2$ and $SrTiO_3$. For $\theta_{GB}^{STO}$ = 24º, the normal state resistivity of the intra-grain bridges was somewhat higher than that of the inter-grain bridges. The semi-logarithmic plots of Figure 5(g,k) proved the resistivity dropped to the detection limit of the voltmeter below the transition.

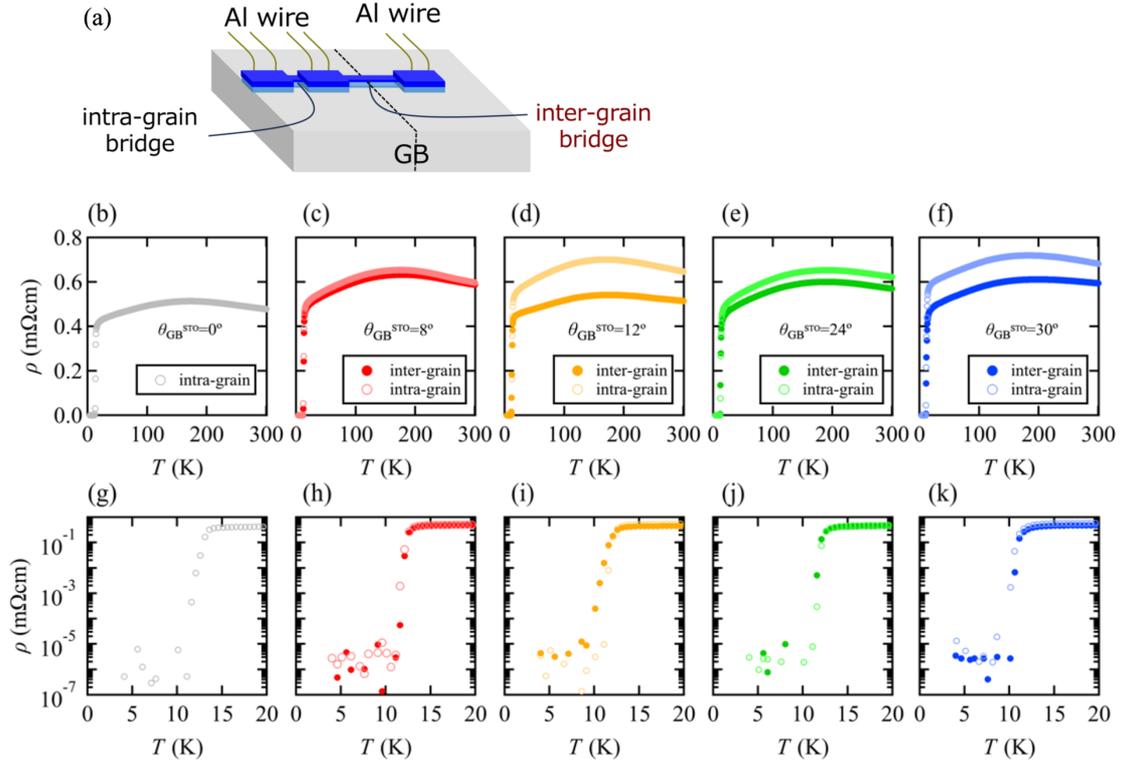

**Figure 5.** The electrical measurements using the intra- and inter-grain bridges is schematized in (a). The resistivity curves of the inter- and intra-grain bridges with various $\theta_{GB}^{STO}$ [(b)~(k)]. The open and solid symbols represent the intra- and inter-grain bridges, respectively. $\theta_{GB}^{STO}$=0º [(b)] refers to the ordinary substrate. Semi-logarithmic plot of (b)~(f) in the vicinity of the transition [(g)~(k)].

Additionally, the transition temperature of the intra-grain and the inter-grain bridges was almost the same for all samples.

Figure 6(a) shows the inter- and intra-grain $J_c$ as a function of $\theta_{GB}^{STO}$ at 4.2 K. For $\theta_{GB}^{FST}$ = 0º, the micro-bridge was fabricated from the film grown on the ordinary SrTiO$_3$ substrate. All bridges showed a $J_c$ of 8×10$^4$ A/cm$^2$ except for the bridge with $\theta_{GB}^{FST}$ = 0º ($J_c$ = 1.6×10$^5$ A/cm$^2$). The reason for higher $J_c$ is that the only $J_c$ component is the *ab*-plane. On the other hand, for $\theta_{GB}^{STO}$ > 0º, the inter-and intra-grain measurements contained two components of $J_c$: along the *c*-axis and the *ab*-plane. However, inter-grain $J_c$ for the films having a $\theta_{GB}^{STO}$ = 24º and 30º was, 8×10$^4$ A/cm$^2$ (Figure 6(b)) although those films had a $\theta_{GB}^{FST}$ close to 0º, Figure 2(b). These results infer that the domain wall structure gave a negative impact on $J_c$. Additionally, the maximum in-plane misorientation was ~3º (Figure 4(d)), which also reduces $J_c$ although the in-plane misorientation angles are less than the critical angle. In fact, the inter-grain $J_c$ of the [001]-tilt GB in Fe(Se,Te) having a misorientation angle of 3º was reduced around 20% relative to the intra-grain $J_c$ [9].

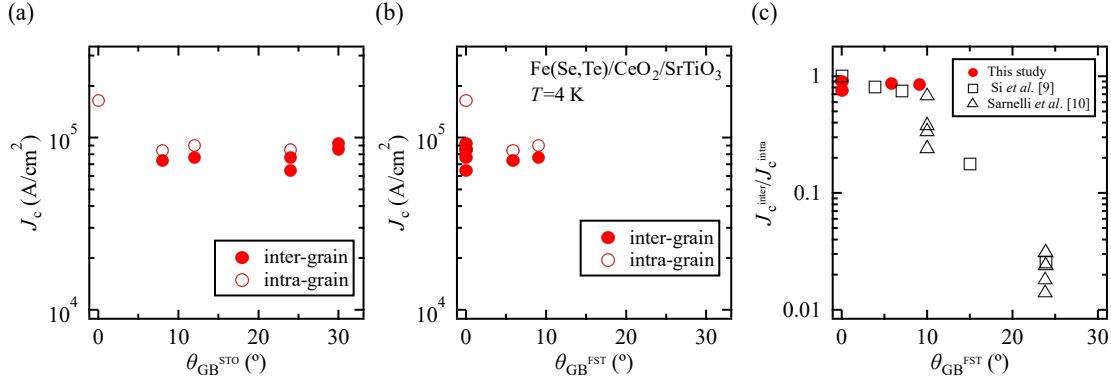

**Figure 6.** $J_c$ of the inter- and intra-grain bridges as a function of $\theta_{GB}^{STO}$ at 4 K. The micro-bridge with $\theta_{GB}^{STO}=0º$ was fabricated from the film grown on the ordinary SrTiO$_3$ substrate. $J_c$ was almost constant around $8\times10^4$ A/cm$^2$ except for the film grown on the ordinary SrTiO$_3$ substrate. (b) Data of (a) replotted as a function of $\theta_{GB}^{FST}$. (c) The $\theta_{GB}^{FST}$ dependence of the normalized $J_c$ for the Fe(Se,Te) bicrystal films measured at 4 K in comparison to data of [001]-tilt GBs (open symbols).

The ratio of inter-grain to intra-grain $J_c$ as a function of $\theta_{GB}^{FST}$ is shown in Figure 6(c). The data for the [001]-tilt GB are also shown for comparison [9,10]. The ratio was almost 1 up to $\theta_{GB}^{FST}$ ~9.5º, which is similar to the [001]-tilt GBs. Hence, the absence of weak-link behavior also for [010]-tilt GBs up to ~9.5º is confirmed in FeSe$_{0.5}$Te$_{0.5}$.

## 4. Conclusion

FeSe$_{0.5}$Te$_{0.5}$ thin films have been grown on CeO$_2$-buffered symmetric [010]-tilt roof-type SrTiO$_3$ bicrystal substrates by pulsed laser deposition. Excess Fe was successfully removed by post-annealing at 200 ºC for 10 min in $p$O$_2$ = 1 Pa. The misorientation angle of the CeO$_2$ buffer layers and FeSe$_{0.5}$Te$_{0.5}$ were different from those of the SrTiO$_3$ bicrystal substrates. The inclined growth of CeO$_2$ can be explained by the geometrical coherency model. For the nominal $\theta_{GB}^{STO}$ = 24º and 30º [010]-tilt SrTiO$_3$ bicrystal substrates, domain wall boundaries rather than grain boundaries were formed in FeSe$_{0.5}$Te$_{0.5}$ due to the epitaxial relation (001)[100] FeSe$_{0.5}$Te$_{0.5}$ ∥ (114)[22$\bar{1}$]CeO$_2$. The inter-grain $J_c$ of the [010]-tilt GBs did not decay for a misorientation angle lower than 9.5°. The current results offer implications for mitigating the weak-link issue in HTS, since CeO$_2$ has been used as common buffer layers for HTS.

## Disclosure statement

No potential conflict of interest was reported by the author(s).

**Funding**

This work was supported by JST CREST Grant Number [JPMJCR18J4]. This work was also partly supported by the Advanced Characterization Platform of the Nanotechnology Platform Japan sponsored by the Ministry of Education, Culture, Sports, Science and Technology (MEXT), Japan.

**References**

[1] Fukuda Y, Kutsukake K, Kojima T, et al. Effects of grain boundary structure and shape of the solid–liquid interface on the growth direction of the grain boundaries in multicrystalline silicon. Cryst Eng Comm. 2022;24(10):1948–1954. doi: 10.1039/D1CE01573G

[2] Dimos D, Chaudhari P, Mannhart J, et al. Orientation dependence of grain-boundary critical currents in $YBa_2Cu_3O_{7-\delta}$ bicrystals. Phys Rev Lett. 1988;61 (2):219–222. doi: 10.1103/PhysRevLett.61.219

[3] Hilgenkamp H, Mannhart J. Grain boundaries in high-$T_c$ superconductors. Rev Mod Phys. 2002;74 (2):485–549. doi: 10.1103/RevModPhys.74.485

[4] Iida K, Hänisch J, Yamamoto A. Grain boundary characteristics of Fe-based superconductors. Supercond Sci Technol. 2020;33(4):043001. doi: 10.1088/1361-6668/ab73ef

[5] Held R, Schneider CW, Mannhart J, et al. Low-angle grain boundaries in $YBa_2Cu_3O_{7-\delta}$ with high critical current densities. Phys Rev B. 2009;79(1):014515. doi: 10.1103/PhysRevB.79.014515

[6] Goyal A, Field DP, Held R, et al. Grain boundary networks in high-performance, heteroepitaxial, YBCO films on polycrystalline, cube-textured metals. Philos Mag Lett. 2011;91(4):246–255. doi: 10.1080/09500839.2010.548345

[7] Zhu Y, Li Q, Tsay YN, et al. Structural origin of misorientation-independent superconducting behavior at [001] twist boundaries in $Bi_2Sr_2CaCu_2O_{8+\delta}$. Phys Rev B. 1998;57(14):8601–8608. doi: 10.1103/PhysRevB.57.8601

[8] Katase T, Ishimaru Y, Tsukamoto A, et al. Advantageous grain boundaries in iron-pnictide superconductors. Nat Commun. 2011;2(1):409. doi: 10.1038/ncomms1419

[9] Si W, Zhang C, Shi X, et al. Grain boundary junctions of $FeSe_{0.5}Te_{0.5}$ thin films on SrTiO3 bi-crystal substrates. Appl Phys Lett. 2015;106(3):032602. doi: 10.1063/1.4906429

[10] Sarnelli E, Nappi C, Camerlingo C, et al. Properties of Fe(Se, Te) bicrystal grain boundary junctions, Squids, and Nanostrips. IEEE Trans Appl Supercond. 2017;27:7400104. doi: 10.1109/TASC.2016.2636248

[11] Omura T, Matsumoto T, Hatano T, et al. Fabrication of grain boundary junctions using NdFeAs (O, F) superconducting thin films. J Phys Conf Ser. 2018;1054:012024. doi: 10.1088/1742-6596/1054/1/012024

[12] Iida K, Omura T, Matsumoto T, et al. Grain boundary characteristics of oxypnictide NdFeAs(O,F) superconductors. Supercond Sci Technol. 2019;32 (7):074003. doi: 10.1088/1361-6668/ab1660

[13] Liu TJ, Ke X, Qian B, et al. Charge-carrier localization induced by excess Fe in the superconductor $Fe_{1+y}Te_{1-x}Se_x$. Phys Rev B. 2009;80(17):174509. doi: 10.1103/PhysRevB.80.174509

[14] Tokuta S, Hasegawa Y, Shimada Y, et al. Enhanced critical current density in K-doped Ba122 polycrystalline bulk superconductors via fast densification. iScience. 2022;25(4):103992. doi: 10.1016/j.isci.2022.103992

[15] Tokuta S, Yamamoto A. Enhanced upper critical field in Co-doped Ba122 superconductors by lattice defect tuning. APL Mater. 2019;7(11):111107. doi: 10.1063/1.5098057

[16] Si W, Han S, Shi X, et al. High current superconductivity in $FeSe_{0.5}Te_{0.5}$-coated conductors at 30 tesla. Nat Commun. 2013;4(1):1347. doi: 10.1038/ncomms2337

[17] Ozaki T, Wu L, Zhang C, et al. A route for a strong increase of critical current in nanostrained iron-based superconductors. Nat Commun. 2016;7(1):13036. doi: 10.1038/ncomms13036

[18] Zhang C, Si W, Li Q. Doubling the critical current density in superconducting $FeSe_{0.5}Te_{0.5}$ thin films by low temperature oxygen annealing. Appl Phys Lett. 2016;109(20):202601. doi: 10.1063/1.4967879

[19] Sun Y, Shi Z, Tamegai T. Review of annealing effects and superconductivity in $Fe_{1+y}Te_{1-x}Se_x$ superconductors. Supercond Sci Technol. 2019;32(10):103001. doi: 10.1088/1361-6668/ab30c2

[20] Rauch EF, Portillo J, Nicolopoulos S, et al. Automated nanocrystal orientation and phase mapping in the transmission electron microscope on the basis of precession electron diffraction. Z fürKristallogr. 2010;225(2–3):103–109. doi: 10.1524/zkri.2010.1205

[21] Guo Z, Gao H, Kondo K, et al. Nanoscale texture and microstructure in a NdFeAs(O,F)/IBAD-MgO superconducting thin film with superior critical current

properties. ACS Appl Electron Mater. 2021;3(7):3158–3166. doi: 10.1021/acsaelm.1c00364

[22] Zhang N, Liu C, Zhao J-L, *et al*. Observation of selective surface element substitution in FeTe$_{0.5}$Se$_{0.5}$ superconductor thin film exposed to ambient air by synchrotron radiation spectroscopy. Chin Phys B. 2016;25(9):097402. doi: 10.1088/1674-1056/25/9/097402

[23] Mukasa K, Matsuura K, Qiu M, *et al*. High-pressure phase diagrams of FeSe$_{1-x}$Te$_x$: correlation between suppressed nematicity and enhanced superconductivity. Nat Commun. 2021;12(1):381. doi: 10.1038/s41467-020-20621-2

[24] Popov R, Ackermann K, Rijckaert H, *et al*. Effect of oxygenation process on flux pinning in pristine and BaHfO3 nanocomposite GdBa2Cu3O7 superconducting thin films. J Phys: Conf Ser. 2020;1559(1):012038. doi: 10.1088/1742-6596/1559/1/012038

[25] Nagai H. Structure of vapor-deposited Ga$_x$In$_{1-x}$ as crystals. J Appl Phys. 1974;45(9):3789–3794. doi: 10.1063/1.1663861

[26] Ayers JE, Ghandhi SK, Schowalter LJ. Crystallographic tilting of heteroepitaxial layers. J Cry Growth. 1991;113 (3–4):430–440. doi: 10.1016/0022-0248(91)90077-I

[27] Budai JD, Yang W, Tamura N, *et al*. X-ray microdiffraction study of growth modes and crystallographic tilts in oxide films on metal substrates. Nature Mater. 2003;2(7):487–492. doi: 10.1038/nmat916

[28] Bryja H, Hühne R, Iida K, *et al*. Deposition and properties of Fe(Se,Te) thin films on vicinal CaF$_2$ substrates. Supercond Sci Technol. 2017;30 (11):115008. doi: 10.1088/1361-6668/aa8421

[29] Kaneko S, Akiyama K, Ito T, *et al*. Large lattice misfit on epitaxial thin film: coincidence site lattice expanded on polar coordinate system. Jpn J Appl Phys. 2010;49 (8S1):08JE02. doi: 10.1143/JJAP.49.08JE02

[30] Lucid AK, Plunkett AC, Watson GW. Predicting the structure of grain boundaries in fluorite-structured materials. Johnson Matthey Technol Rev. 2019;63 (4):247–254. doi: 10.1595/205651319X15598975874659

[31] Feng B, Sugiyama I, Hojo H, *et al*. Atomic structures and oxygen dynamics of CeO$_2$ grain boundaries. Sci Rep. 2016;6(1):20288. doi: 10.1038/srep20288

**Supplementary information on "Structural analysis and transport properties of the [010]-tilt grain boundaries in Fe(Se,Te)"**

Kazumasa Iida[1,8], Yoshihiro Yamauchi[2], Takafumi Hatano[2,8], Kai Walter[3], Bernhard Holzapfel[3], Jens Hänisch[3], Zimeng Guo[4,8], Hongye Gao[5], Haoshan Shi[6], Shinnosuke Tokuta[7,8], Satoshi Hata[4,5,6,8], Akiyasu Yamamoto[7,8], Hiroshi Ikuta[2,9]

1 College of Industrial Technology, Nihon University, 1-2-1 Izumi-cho, Narashino, Chiba 275-8575, Japan;

2 Department of Materials Physics, Nagoya University, Furo-cho, Nagoya 464-8603, Japan;

3 Institute for Technical Physics, Karlsruhe Institute of Technology, Hermann-von-Helmholtz-Platz 1, Eggenstein-Leopoldshafen, 76344, Germany;

4 Department of Advanced Materials Science and Engineering, Kyushu University, Kasuga, Fukuoka 816-8580, Japan;

5 The Ultramicroscopy Research Center, Kyushu University, Motooka, Fukuoka 819-0395, Japan;

6 Interdisciplinary Graduate School of Engineering Sciences, Kyushu University, Kasuga, Fukuoka 816-8580, Japan;

7 Department of Applied Physics, Tokyo University of Agriculture and Technology, Koganei, Tokyo 184-8588, Japan;

8 JST CREST, Kawaguchi, Saitama 332-0012, Japan;

9 Research Center for Crystalline Materials Engineering, Nagoya University, Furo-cho, Nagoya 464-8603, Japan

CONTACT Kazumasa Iida: iida.kazumasa@nihon-u.ac.jp

## 1. X-ray reflectivity measurement for CeO₂ grown on SrTiO₃(001)

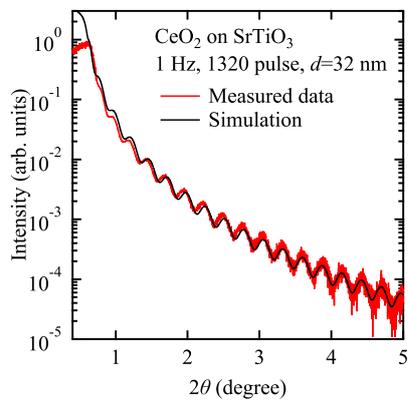

**Figure S1|** X-ray reflectivity measurement confirmed that the thickness of CeO₂ is 32 nm, which is almost identical to the one measured from the cross-sectional ADF-STEM image [fig. 3(a)].

**2. The 101 reflection of $\phi$ scans and the 00$l$ rocking curves of FeSe$_{0.5}$Te$_{0.5}$ grown on the CeO$_2$-buffered SrTiO$_3$ ordinary substrate and bicrystal substrate with $\theta_{GB}^{STO}$=30°.**

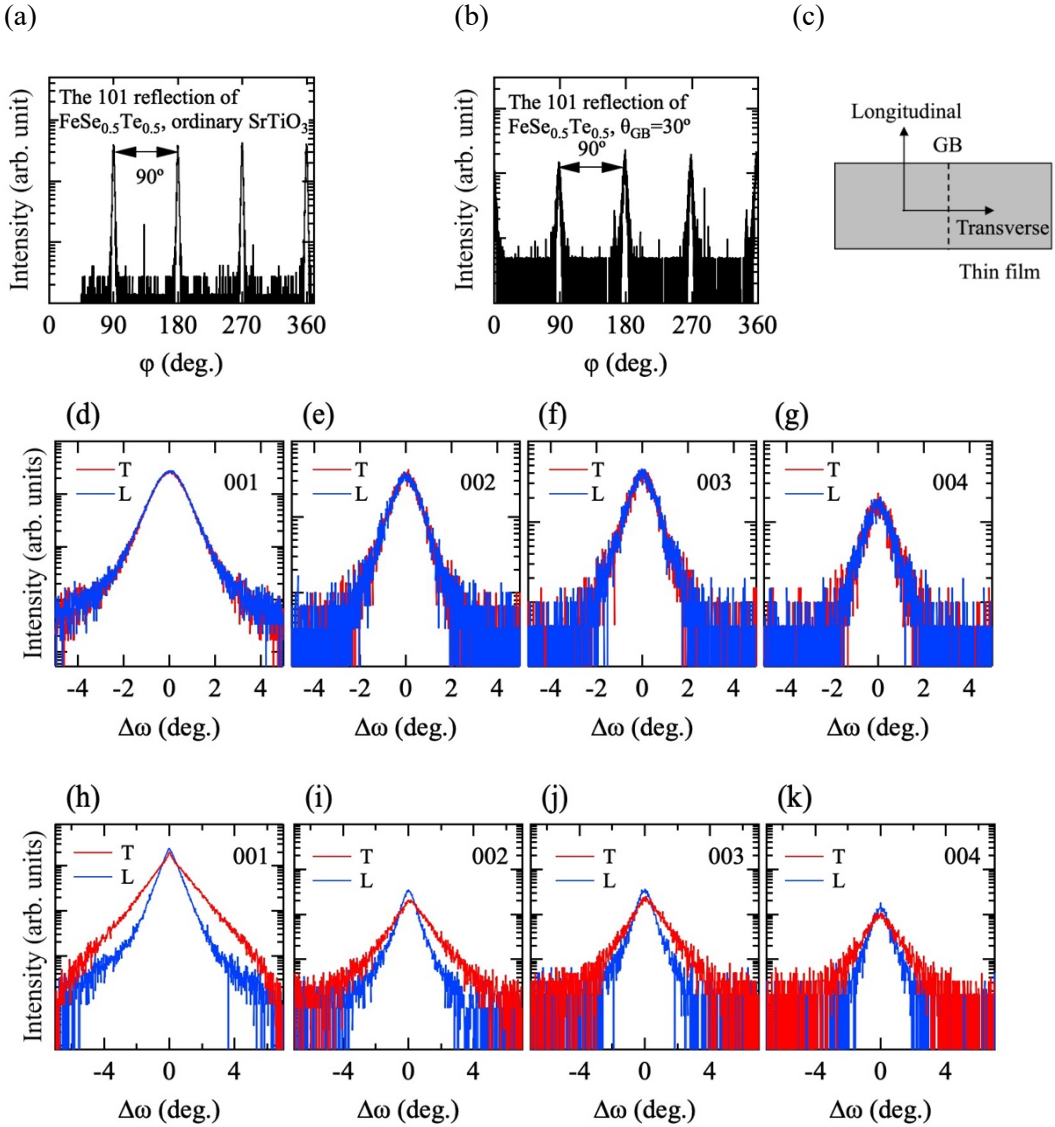

**Figure S2**| (a) The $\phi$ scans of the 101 reflection of FeSe$_{0.5}$Te$_{0.5}$ grown on CeO$_2$-buffered ordinary SrTiO$_3$ substrate and (b) bicrystal substrate with $\theta_{GB}^{STO}$=30°. (c) The schematic illustration of the X-ray scan direction for the 00$l$ rocking curves. (d)-(g) The 00$l$ ($l$=1, 2, 3 and 4) rocking curves of FeSe$_{0.5}$Te$_{0.5}$ grown on CeO$_2$-buffered ordinary SrTiO$_3$ substrate. "T" and "L" denote the transverse and longitudinal directions of the X-ray scans, shown in (c). As expected, no difference in FWHM were observed. On the other hand, for the film grown on bicrystal substrate ($\theta_{GB}^{STO}$=30°), the FWHM for "T" (i.e., perpendicular to the GB) are larger than those for "L" [(h)~(k)].

Table S1 shows the FWHM exhibited in fig. S2(a) and (b). The $\Delta\phi$ of the FeSe$_{0.5}$Te$_{0.5}$ film on single crystal SrTiO$_3$ ($\theta_{GB}^{STO} = 0°$) is smaller than of the film on bicrystal substrate. Table S2 summarizes the crystalline quality of the out-of-plane direction of the FeSe$_{0.5}$Te$_{0.5}$ films on single crystal and bicrystal substrates. The FWHM of FeSe$_{0.5}$Te$_{0.5}$ on ordinary SrTiO$_3$ are almost the same values regardless of the scan directions. On the other hand, for FeSe$_{0.5}$Te$_{0.5}$ on bicrystal substrate, FWHM shows a strong directional dependence: FWHM of "T"-direction are always larger than those of "L"-direction.

**Table S1|** The FWHM ($\Delta\phi$) of FeSe$_{0.5}$Te$_{0.5}$ grown on the CeO$_2$-buffered ordinary SrTiO$_3$ substrate and bicrystal SrTiO$_3$ substrate ($\theta_{GB}^{STO}$=30°).

| FeSe$_{0.5}$Te$_{0.5}$ on | $\Delta\phi_{101}$ (°) | $\Delta\phi_{011}$ (°) | $\Delta\phi_{\bar{1}01}$ (°) | $\Delta\phi_{0\bar{1}1}$ (°) |
|---|---|---|---|---|
| ordinary SrTiO$_3$ | 2.20 | 2.24 | 2.15 | 2.28 |
| bicrystal SrTiO$_3$ ($\theta_{GB}^{STO}$=30°) | 3.82 | 3.95 | 3.82 | 4.07 |

**Table S2|** The FWHM ($\Delta\omega$) of FeSe$_{0.5}$Te$_{0.5}$ grown on the CeO$_2$-buffered ordinary SrTiO$_3$ substrate and bicrystal SrTiO$_3$ substrate ($\theta_{GB}^{STO}$=30°).

| FeSe$_{0.5}$Te$_{0.5}$ on | direction | $\Delta\omega_{001}$ (°) | $\Delta\omega_{002}$ (°) | $\Delta\omega_{003}$ (°) | $\Delta\omega_{004}$ (°) |
|---|---|---|---|---|---|
| ordinary SrTiO$_3$ | T | 1.26 | 1.15 | 1.15 | 1.13 |
| | L | 1.25 | 1.16 | 1.13 | 1.07 |
| bicrystal SrTiO$_3$ ($\theta_{GB}^{STO}$=30°) | T | 2.73 | 2.27 | 1.64 | 1.59 |
| | L | 1.19 | 0.83 | 0.87 | 0.91 |

## 2. Vicinal angle evaluated from the geometrical coherency model

To evaluate vicinal angles of FeSe$_{0.5}$Te$_{0.5}$ grown on off-cut CaF$_2$ substrates, the following equation is employed:

$$\Delta\theta_{vic}^{FST} = \left| \tan^{-1}\left( \frac{d_{CaF2} - d_{FST}}{d_{CaF2}}\tan\theta_{vic}^{CaF2} \right) \right| \quad (S1)$$

with

$$\theta_{vic}^{FST} = \Delta\theta_{vic}^{FST} + \theta_{vic}^{CaF2} \quad (S2)$$

where $\theta_{vic}^{FST}$ is the vicinal angle of FeSe$_{0.5}$Te$_{0.5}$, $d_{CaF2}$ is the lattice parameter of CaF$_2$ (5.462 Å), $d_{FST}$ is the $c$-axis length of FeSe$_{0.5}$Te$_{0.5}$ (5.96 Å) [S1] and $\theta_{vic}^{CaF2}$ is the vicinal angle of the CaF$_2$ substrates. Due to $d_{FST} > d_{CaF2}$, the direction of the tilt of [001] FeSe$_{0.5}$Te$_{0.5}$ from [001] CaF$_2$ is away from the substrate normal. The FeSe$_{0.5}$Te$_{0.5}$ thin films were grown on off-cut CaF$_2$ substrates at 260 °C and 400 °C, respectively [S2]. Figure S3 shows the measured vicinal angle of FeSe$_{0.5}$Te$_{0.5}$ by X-ray diffraction as a function of $\theta_{vic}^{CaF2}$. The dashed lines are calculation from equations S1 and S2. As can be seen, the vicinal angles of FeSe$_{0.5}$Te$_{0.5}$ grown at 260 °C deviate from the calculation, whereas they lie on the calculated lines for the films grown at 400 °C. These results suggest that the inclined growth mechanism based on the geometrical coherence model is operative at high growth temperature, but not at low growth temperature.

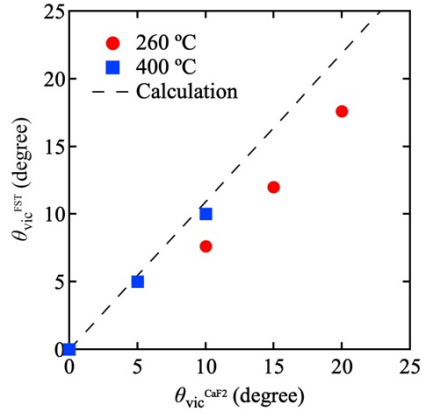

**Figure S3|** The vicinal angle of FeSe$_{0.5}$Te$_{0.5}$ on off-cut CaF$_2$ substrates grown at 260 °C and 400 °C as a function of $\theta_{vic}^{CaF2}$.

## 4. Cross-sectional TEM image of FeSe$_{0.5}$Te$_{0.5}$ on CeO$_2$-buffered ordinary SrTiO$_3$ substrate

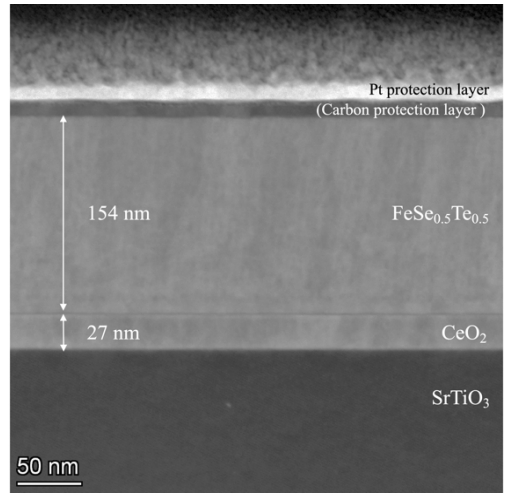

**Figure S4|** The cross-sectional view of FeSe$_{0.5}$Te$_{0.5}$ grown on the CeO$_2$-buffered SrTiO$_3$ substrate ($\theta_{GB}^{STO} = 0°$) obtained by ADF-STEM.

## Reference

[S1] Musaka K, Matsuura K, Qin M, Saito M, Sugiura Y, Ishida K, Otani M, Oishi Y, Mizukami Y, Hashimoto K, Gouchi J, Kumai R, Uwatoko Y, Shibauchi T. High-pressure phase diagrams of FeSe$_{1-x}$Te$_x$: correlation between suppressed nematicity and enhanced superconductivity. Nat. Commun. (2021);12: 381. doi: 10.1038/s41467-020-20621-2

[S2] Bryja H, Hühne R, Iida K, Molata S, Sala A, Putti M, Schultz L, Nielsch K, Hänisch J.

Deposition and properties of Fe(Se,Te) thin films on vicinal CaF$_2$ substrates. Supercond. Sci. Technol. 2017; 30: 115008. doi: 10.1088/1361-6668/aa8421